# Joint Opportunistic Scheduling in Multi-Cellular Systems

Sugumar Murugesan, Philip Schniter

*The Ohio State University*
*Department of Electrical and Computer Engineering*
*2015 Neil Avenue, Columbus, Ohio 43210*

April 2009

**Abstract**

We address the problem of multiuser scheduling with partial channel information in a multi-cell environment. The scheduling problem is formulated jointly with the ARQ based channel learning process and the intercell interference mitigating cell breathing protocol. The optimal joint scheduling policy under various system constraints is established. The general problem is posed as a generalized Restless Multiarmed Bandit process and the notion of indexability is studied. We conjecture, with numerical support, that the multicell multiuser scheduling problem is indexable and obtain a partial structure of the index policy.

# 1 Introduction

Cellular wireless networks, typically characterized by a central controller (base station) coordinating downlink and uplink transmissions to and from users within a cell, has been a popular model among wireless network designers. A well known application deploying the cellular network model is the cellular wireless telephony [1]. In recent years, there has been a tremendous increase in the demand for high data rates in these systems. The need for spectrally efficient communication strategies is thus on a steady rise. This is particularly serious in cellular systems that are prone to scalability issues. One such strategy is the *opportunistic multiuser scheduling* proposed by Knopp and Humblet, [2], in a single cell environment. Opportunistic multiuser scheduling can be defined as *allocating the system resources to the user(s) experiencing the most favorable channel conditions at the moment*. It is particularly well suited to the cellular environment thanks to the presence of the base station, a central coordinating authority.

Opportunistic scheduling has since been studied extensively under various scenarios [3, 4, 5, 6, 7]. For a general treatment on the topic with minimal physical layer assumptions, see [8]. It is understandable that the availability of the channel state information at the scheduler is crucial for the success of the opportunistic scheduling schemes. A vast majority of the literature on this topic make the unrealistic assumption of readily available channel state information at the scheduler. In reality, however, a non-trivial amount of resource must be spent in gathering the information on the channel state. This leads us to the following critical question: *Are there efficient joint channel acquisition - multiuser scheduling schemes for a cellular system ?* For single cell systems, this issue has recently been addressed in independent works [9] and [10] in the contexts of multiuser downlink and cognitive radio, respectively. These works exploit the memory inherent in the fading channels along with the ARQ feedback used for error control purposes for opportunistic channel aware scheduling. They model the fading channel with memory by a Gilbert Elliott channel ([11, 12, 13, 14, 15, 16]) and identify the effect of channel scheduling decisions on the channel information acquisition process and vice versa. By formulating the joint optimization problem as a dynamic program, they show that a greedy policy that maximizes the immediate reward (reward defined to reflect data rate in [9]) is optimal. The policy is also shown to be remarkably simple to implement. Notice that there is no additional overhead incurred by the channel acquisition process. These works underline the potential for tapping into existing resources in the system (in this case, the channel memory and the ARQ mechanism already in place for error control) for low overhead channel aware scheduling.

Rarely in realistic scenarios do we encounter a single cell cellular system. The very idea of cells was conceived with an intention to control wireless communication between users spread over a large geographic area by splitting it into small manageable cells. Thus transmission in a cell interferes with the transmissions in the adjacent cells. It follows that the channel state of any user in a cell is a function of the transmissions and schedule decisions in the adjacent cells, effectively imparting a convolved dependence between the scheduling choices in neighboring cells. We now face the following question: *How does the easily implementable, low overhead, ARQ based joint channel acquisition - multiuser scheduling scheme extend to the multi-cell environment ?*

We address this problem by following a two layered approach: A well established inter-cell interference (ICI) control mechanism is adopted and assumed to be in place. On top of this layer we optimize the ARQ based multiuser scheduling scheme across the cells. We now proceed to introduce our choice of the ICI mechanism after a short literature survey on the topic.

Traditionally, ICI is controlled by staggering the transmissions in adjacent cells across orthogonal frequency bands and reusing these bands in geographically far-apart cells. This is the well known frequency reuse based ICI control mechanism [1] that is prevalent in narrow band systems, such as the GSM. Other ICI control mechanisms have also been studied. In [17], a capture division packet access (CDPA) mechanism is proposed. Here, users are allowed to transmit on the same carrier in adjacent cells, i.e., no frequency reuse based ICI control is deployed. The effect of interference is quantified by a capture probability defined as the likelihood of successful transmission under ICI. Upon collision, a retransmission is performed. The authors demonstrate that CDPA outperforms traditional TDMA based strategies under certain operating conditions.

Notice that, in the preceding scenario, users at the periphery of the cell suffer from low signal to interference ratio and hence low capture probability compared to the users near the base station. This is the classic *near-far* effect. If this is not addressed properly, under QoS requirements on fairness across users, the far users will act as a bottleneck thus bringing down the overall system utility. Taking note of this crucial phenomenon, the authors in [18] propose a novel reduced power channel reuse (RPCR) scheme that aims to equalize the capture probabilities of the near and far users. By formally classifying the users into two groups: near and far (based on a generic "distance" metric that need not be a function of the geometric distance), RPCR works as follows: If, in a carrier, a far user is scheduled in a cell, the power of transmission in the same carrier in the adjacent cell is deliberately reduced. This power reduction naturally limits transmission to near users in the adjacent cell. Thus, at any time, in any carrier, cell 1 and cell 2 transmits to users belonging to complementary groups with different power levels (full power for the far user). With the availability of multiple carriers, in a cell, carriers are classified as primary and secondary with far users assigned to the primary carriers (at full power level) and near users to the secondary carriers (at reduced power levels). This primary/secondary classification is reversed in the adjacent cell. With this arrangement, the two cells need not coordinate their transmissions to maintain the main idea of the RPCR. The authors formulate and study the optimal channel selection policy that assigns users to the near/far groups. They show that the RPCR scheme is superior in performance to other ICI control mechanisms in terms of sum throughput under uniform fairness constraints.

In [19], the authors take a fundamental, information theoretic approach towards ICI control. They show that varying power across carriers in a complementary fashion across adjacent cells with far users assigned to carriers with higher power and vice versa improves the overall capacity region and hence the spectral efficiency of a two-cell system. For reasonable fairness constraints, the spectral efficiency was shown to be better in comparison to the traditional frequency reuse based ICI control. This is consistent with the observation made in [18] with respect to the RPCR scheme. In [19], the authors also consider a multi-cell system with wideband mobiles. Here the mobile carriers are assumed to communicate over the entire available spectrum, a possibility in wideband systems. Thus, with only one carrier available, the luxury of varying power across carriers is absent. For this scenario, the authors propose a *cell-breathing* scheme where, in a cell, the power allocated to the carrier varies rhythmically across time, in a fashion complementary to the adjacent cells. Thus, at no time, the adjacent cells transmit at the same power (and hence to the same group users). This rhythmic power variation across time resembles a breathing pattern, hence the name cell breathing. Notice that the main idea of equalizing the capture probabilities of the near and far users and hence the utility gains associated with the approach are retained in the cell breathing technique. This is demonstrated in [19]. The cell-breathing protocol for wideband systems was studied further in [20]. Here the authors obtain achievable rate regions associated with the cell breathing strategy when it is deployed alone and in combination with several precoding schemes.

Encouraged by the positive results associated with the RPCR and the cell breathing techniques, we adopt cell breathing as our ICI control mechanism. Note that we are assuming a wideband system with only one carrier available. We will later show that, with regard to our analysis in this work, this is a general case than the system with multiple carriers. If the channel of the users are time-variant, it is readily seen that, without violating the cell breathing protocol, the performance of the system can be improved by (joint) opportunistic multiuser scheduling with coordination across cells. We address this joint opportunistic scheduling problem in a two-cell system[1] in this work. By demonstrating that the channel can still be modeled by i.i.d two-state Markov chains, like in the single cell case, we study the ARQ based joint opportunistic scheduling scheme in the two-cell system, under various scenarios: (a) When the cooperation between the cells is not symmetric. (b) When there are restrictions on the breathing pattern. Here the breathing pattern refers to the time-sequence of near-far-near... users scheduled across time. A simple illustration of the breathing pattern with and without opportunistic scheduling is available in Fig. 1. (c) When there is complete cooperation with no restriction on the breathing pattern. In the last scenario, a direct analysis of the problem appears intractable. We therefore establish a link between our problem and the *Restless Multiarmed Bandit (RMAB) processes*. We introduce the notion of indexability from the RMAB theory and perform an indexability analysis for the current system. We claim, with numerical support, that the scheduling problem at hand is in fact indexable and partially characterize the link between the index based policy and the greedy policy.

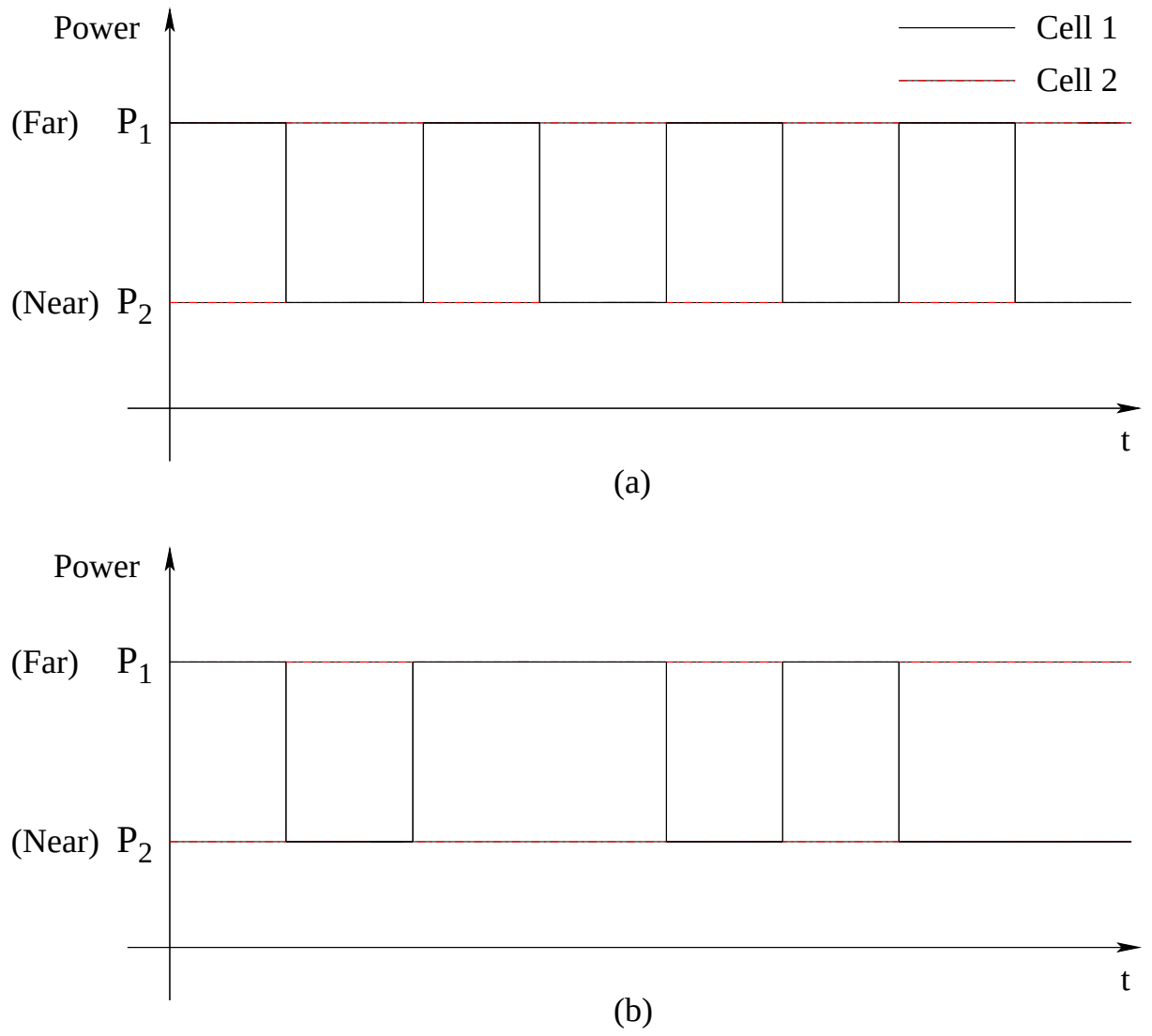

Figure 1: Breathing Pattern Illustration: (a) The breathing is rhythmic without joint opportunistic scheduling (b) The breathing pattern is influenced by joint opportunistic scheduling.

The report is organized as follows. The problem setup is described in Section 2 followed by a study of the optimal ARQ based scheduling policy under different system requirements in Section 3. We establish the connection between the scheduling problem and RMAB processes along with an overview of the notion of indexability in Section 4. We perform an indexability analysis of the scheduling problem in Section 5. In Section 6, we partially characterize the link between the index policy and the greedy policy.

[1]Extension to multi-cell system is discussed in the next section

# 2 Problem Setup

## 2.1 Channel Model

Consider a two-cell system. Consistent with [18, 19], within each cell, we cluster users into near and far users. We use geometric distance between the users and their respective base stations as the metric for this classification. Denote by $n_i, f_i$, the set of near and far users, respectively in cell $i \in \{1, 2\}$. An user in a group is denoted by the label of the group for notational simplicity. Let the distance between base station $i$ and user $j$ (in any cell) be given by $d_{ij}$. By way of the two level clustering we assume, $d_{if_i}$ is the same for all far users in cell $i$. Likewise $d_{in_i}$ is uniform for near users in cell $i$. Let $N$ be the number of near users in the cells and $F$ the number of far users. Denote the normalized (with respect to attenuation loss) fading coefficient of a link between base station $i$ and user $j$ (in any cell) as $h_{ij}$. We assume $h_{ij}$ are i.i.d distributed. Consider cell 1 as the primary cell and cell 2 the interfering cell. If a far user $f$ is[2] served in the primary cell with power $P_f$ and if the interfering base station is transmitting at power $P_{I_f}$ ($I_f$ indicates interference to the far user in the primary cell) then the SINR at this user is given as below.

$$\mathrm{SINR}_f = \frac{\frac{P_f}{d_{1f}^{\alpha}}|h_{1f}|^2}{N_0 + \frac{P_{I_f}}{d_{2f}^{\alpha}}|h_{2f}|^2} \tag{1}$$

where $N_0$ indicates the variance of the additive noise. We have used the attenuation model from [21] with $\alpha \geq 2$ being the attenuation coefficient. Likewise, if a near user is served in the primary cell with the interfering base station power being $P_{I_n}$, the SINR is given by

$$\mathrm{SINR}_n = \frac{\frac{P_n}{d_{1n}^{\alpha}}|h_{1n}|^2}{N_0 + \frac{P_{I_n}}{d_{2n}^{\alpha}}|h_{2n}|^2}. \tag{2}$$

An illustration of these two scenarios is provided in Fig. 2.

Consistent with the two level clustering we assume, the base stations are allowed to choose one of two power levels, i.e., $P_f, P_n, P_{I_f}, P_{I_n} \in \{P_1, P_2\}$ with $P_2 < P_1$. By observation, since $d_{1f} > d_{1n}$, the avarage SINR of the far and near users can be equalized if $P_{I_f} < P_{I_n}$ and $P_f > P_n$. This, along with the constraint on the alphabet size of the power levels, leads to the cell breathing rule [18, 19, 20]: *A far user is served with power $P_1$ and a near user with power $P_2 < P_1$. Whenever a far user is scheduled in a cell, a near user is scheduled in the adjacent cell and vice versa.*

Since the links between the base stations and users $h_{ij}$ are i.i.d, with the SINR values equalized under cell breathing, we have the following: $SINR_{f_1}, SINR_{n_1}, SINR_{f_2}, SINR_{n_2}$ are i.i.d. The fading coefficients were assumed to evolve with memory in [9], leading to the GE model. Retaining this assumption on the fading coefficients and assuming a threshold based decoding rule as below: SINR $\geq \gamma$[3] $\Rightarrow$ decoding success and SINR $< \gamma \Rightarrow$ decoding

[2] We have dropped the suffix 1 as the context is clear.

[3] The value of the threshold being a function of the application and the decoder in use

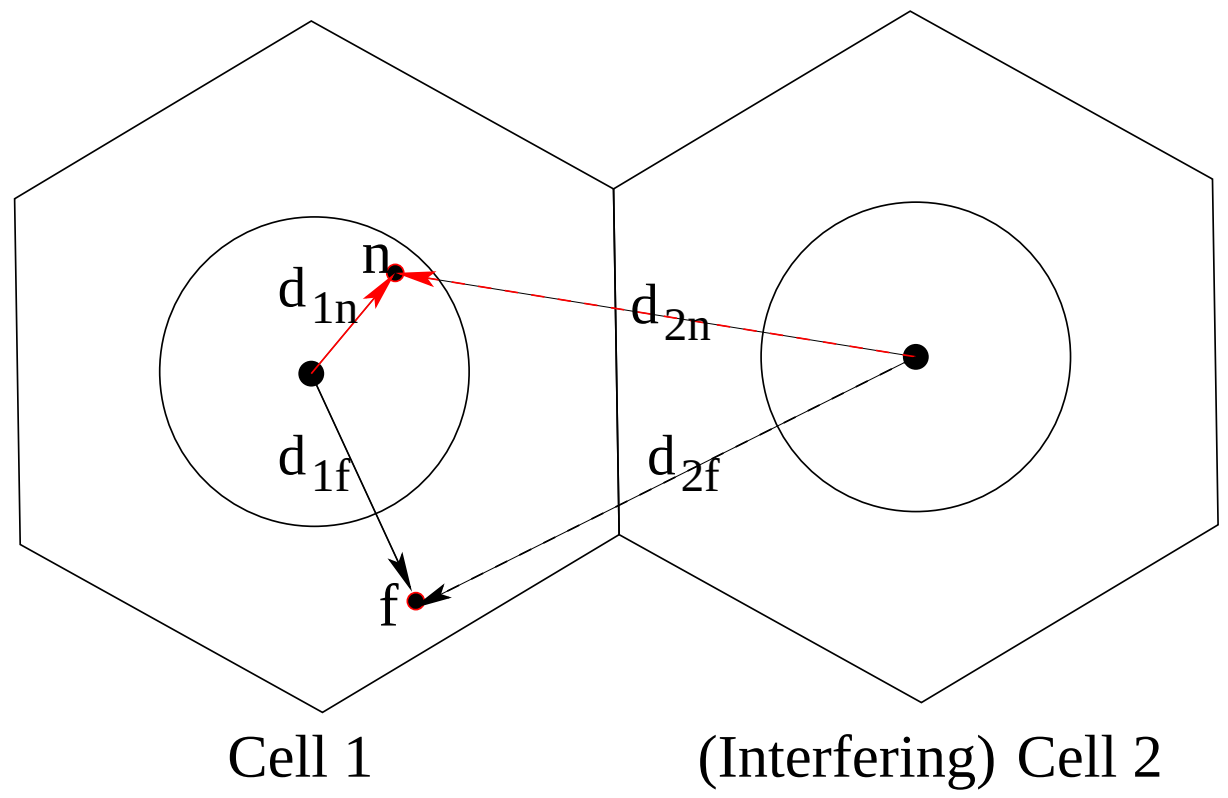

Figure 2: Illustration showing transmissions and interference caused when a far user and a near user are served (at different times).

failure, we retain the GE model for the channels of the users. Specifically, the channel of each user is modeled by an i.i.d two-state Markov chain, with the ON state allowing the successful transmission of a fixed length packet. The channel of each user remains fixed through a time slot and evolves into another state in the next slot according to the Markov chain statistics. The time slots of all users are synchronized. The two-state Markov channel is characterized by a $2 \times 2$ probability transition matrix

$$P = \begin{bmatrix} p & q \\ r & s \end{bmatrix}, \tag{3}$$

where

- $p :=$ prob(channel is ON in the current slot | channel was ON in the previous slot)
- $q := 1 - p$
- $r :=$ prob(channel is ON in the current slot | channel was OFF in the previous slot)
- $s := 1 - r$.

Throughout this work, we consider positively correlated (in time) channels, i.e., $p \geq r$.

It is worth noting that the scheduling analysis for the two-cell system readily extends to a multi-cell configuration with the use of six-directional antennae ([1]) at the base stations. Each cell can now be divided into six regions and the joint scheduling analysis in this work can be applied in each region independently. This is illustrated in Fig. 3.

## 2.2 Scheduling Problem

The base stations are the central controllers that control the transmission to the users in each control interval within their respective cells. In each control interval, the base stations do not know the *exact* channel state of the users and must schedule the transmission of the head of line packet of exactly one user (a data queue is maintained for each user to collect the data meant for that user) each, while maintaining the cell breathing protocol. In any cell, if a far user is scheduled, transmission takes place at full power $P_1$. For a near user the lower power $P_2 < P_1$ is used, in accordance with cell breathing. A traditional ARQ based transmission is deployed in each cell. Here, in each cell, at the beginning of a time slot, the head of line packet of the scheduled user is transmitted. If the packet does not go through, i.e., it cannot be decoded by the user (when the channel is in OFF state), a NACK is sent back from the user at the end of the slot, and the packet is retained at the head of the queue for retransmission at a later opportunity. If the packet goes through (ON state), an ACK is sent back and the packet is removed from the queue. The ARQ feedback is assumed to be transmitted over a dedicated error free channel. At the end of the slot, the base stations of the neighboring cells share this ARQ information. Thus each base station has all the channel information that are available to its neighbor. This is crucial in the cell breathing based joint scheduling. Thus, effectively, we may consider the base station pair as a single control entity when it comes to joint scheduling based on ARQ feedbacks. The performance metric that the base stations aim to maximize is the sum throughput of the system. Details are discussed next.

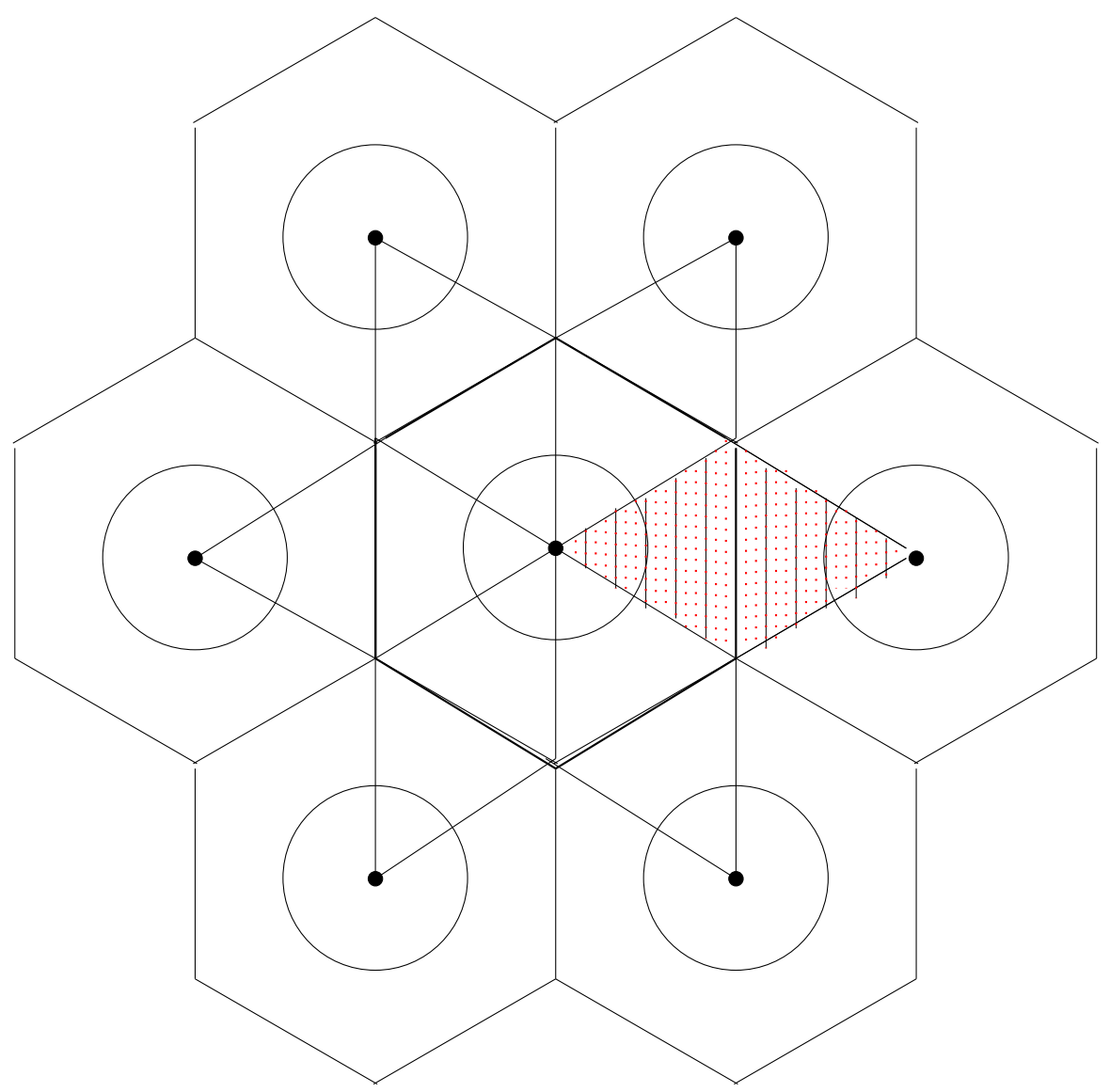

Figure 3: Multi-cell extension: with six directional antennae at the base stations, each cell can be split into six regions and the two-cell joint scheduling can be performed on these regions independently. One such region is highlighted.

## 2.3 Formal Problem Definition

We now introduce the terms/entities that we use in this work, many of which are borrowed from the POMDP [22] literature.

*Control interval $k$*: Each time slot in our problem setup will henceforth be called a control interval. The scheduling process horizon is fixed. A control interval is indexed by $k$ if there are $k-1$ more intervals until the horizon.

*Action* $(\mathsf{a}_k, a_k)$: Indicates the indices of the user pair scheduled in cells 1 and 2 in control interval $k$. With cell breathing in place, we have the following constraint: $(\mathsf{a}_k, a_k) \in \{(n_1, f_2), (f_1, n_2)\}$. We denote this admissible set by $\mathcal{B}$.

*Belief values at the $k^{th}$ control interval*: Denote by $\pi_k^c$, the vector of the belief values (the probability of having an ON state) of the users in cell $c$ at time $k$. Let $F_k^c$ indicate the ARQ feedback received at the end of control interval $k$ in cell $c$. We denote an ACK by 1 and a NACK by 0. The belief values in cell 1 evolve as below:

$$\pi_{k-1}^1(i) = \begin{cases} p, & \text{if } i = \mathsf{a}_k, F_k^1 = 1 \\ r, & \text{if } i = \mathsf{a}_k, F_k^1 = 0 \\ p\pi_k^1(i) + r(1 - \pi_k^1(i), & \text{if } i \neq \mathsf{a}_k. \end{cases} \tag{4}$$

where the first case indicates that, in cell 1, user $i$ is scheduled in control interval $k$ and an ACK feedback was received. Thus, according to the Markov chain statistics, $\pi_{k-1}^1(i) = p$. The second case is explained similarly when a NACK feedback is received. The last case indicates that user $i$ was not scheduled for transmission in control interval $k$ and hence the cell 1 base station must estimate the belief value at the current control interval from that at the previous control interval and the Markov chain statistics. Similar evolution holds for cell 2. It is a well known fact ([22]) that the belief values are sufficient statistics to any information about the channels in the past control intervals, in our case, the scheduling decisions and the ARQ feedbacks from the past. Thus the joint scheduling decision in any control interval can be solely based on the belief values for that interval and not on the past ARQ or schedule information.

*Scheduling Policy* $\mathfrak{A}_k$: A scheduling policy $\mathfrak{A}_k$ in the control interval $k$ is a mapping from the belief values and the control interval index to an action as follows:

$$\mathfrak{A}_k : (\{\pi_k^1, \pi_k^2\}, k) \to (\mathsf{a}_k, a_k) \in \mathcal{B}.$$

*Reward Structure*: In any control interval $k$, in cell $c$, a reward of 1 is accrued when the transmission in cell $c$ is successful, i.e, when $F_k^c = 1$, and no reward is accrued otherwise. The total reward in control interval $k$ is simply the sum of the rewards accrued by cells 1 and 2. Note that this reward structure is defined to be consistent with our performance metric, the sum throughput (to be discussed shortly).

*Net Expected Reward in the control interval $m$, $V_m$*: With the belief vectors, $\pi_m^1, \pi_m^2$, and the scheduling policy, $\{\mathfrak{A}_k\}_{k \leq m}$, fixed, the net expected reward, $V_m$, is the sum of the reward, $R_m(\pi_m^1, \pi_m^2, \mathsf{a}_m, a_m)$, expected in the current control interval $m$ and $\mathrm{E}[V_{m-1}]$, the net reward expected in the future control intervals conditioned on the belief vectors and the scheduling decisions in the current control interval. Formally,

$$\begin{aligned} & V_m(\pi_m^1, \pi_m^2, \{\mathfrak{A}_k\}_{k \leq m}) \\ &= R_m(\pi_m^1, \pi_m^2, \mathsf{a}_m, a_m) + \mathrm{E}[V_{m-1}(\pi_{m-1}^1, \pi_{m-1}^2, \{\mathfrak{A}_k\}_{k \leq m-1}) | \pi_m^1, \pi_m^2, \mathsf{a}_m, a_m], \end{aligned} \tag{5}$$

where the expectation is over the belief vectors $\pi_{m-1}^1, \pi_{m-1}^2$. Since the reward in each control interval in each cell is either 1 or 0, the expected current reward can be written

as

$$R_m(\pi_m, a_m) \;=\; \pi_m^1(\mathsf{a}_m) + \pi_m^2(a_m).$$

*Performance Metric- the Sum Throughput, $\eta_{sum}$*: For a given scheduling policy $\{\mathfrak{A}_k\}_{k\geq 1}$, and initial belief vectors $\pi_I^1, \pi_I^2$, the sum throughput is given by

$$\eta_{\text{sum}}(\{\mathfrak{A}_k\}_{k\geq 1}) \;=\; \lim_{m\to\infty} \frac{V_m(\pi_I^1, \pi_I^2, \{\mathfrak{A}_k\}_{k\geq 1})}{m}. \tag{6}$$

*Optimal Scheduling Policy, $\{\mathfrak{A}_k^*\}_{k\geq 1}$*:

$$\{\mathfrak{A}_k^*\}_{k\geq 1} \;:=\; \arg\max_{\{\mathfrak{A}_k\}_{k\geq 1}} \eta_{\text{sum}}(\{\mathfrak{A}_k\}_{k\geq 1}). \tag{7}$$

Before we analyze scheduling in the original system, we consider a few variants of the system in the next section.

# 3 Optimal Policy for Variants of the System

## 3.1 When the Cooperation between the Cells is Asymmetric

Consider a system where the cell breathing is deployed by the following asymmetric cooperation between the cells: base station 1 schedules transmission to its users without any regard to the decisions in cell 2, while base station 2 chooses the group of the user to be scheduled based on the user group choice of base station 1. Base station 1 is aware of this compromise made by base station 2 and therefore adopts the two state Markov model for the channels of its users, which is valid only under cell breathing. Such an asymmetric cooperation can result in scenarios such as (1) Cell 1 may cover the heart of a city with higher data rate requirements compared to cell 2 that covers the suburbs. (2) The sharing of ARQ feedback information between the adjacent base stations may not be mutual due to a link failure between the base stations. (3) In the context of game theory, when base station 1 is a selfish player and base station 2 is a rule-abiding player.

Consider cell 1 under the asymmetric cooperation scenario. Since base station 1 does not make any effort in maintaining cell breathing, the multiuser scheduling problem in cell 1 becomes the same as the multiuser scheduling in an isolated cell. For an isolated Markov modeled downlink with $N$ users, the greedy policy that maximizes the immediate reward is shown to be optimal in [9] (for $N \leq 3$ users) and [10] (for any $N$). Thus base station 1, under asymmetric cooperation, implements the greedy policy within its cell.

We now proceed to study the optimal scheduling policy in cell 2. Fix a realization of the channel states of the users in cell 1 from time $t > 0$ until the horizon. With a fixed scheduling policy in cell 1 (in this case, the greedy policy), we can define a sequence of time instants $\{t_{\mathbf{n}}, t_{\mathbf{n}-1}, \ldots t_1\}$ with $t \geq t_{\mathbf{n}} \geq t_{\mathbf{n}-1}, \ldots t_1 \geq 1$, where a near user is scheduled in cell 1. Note that $\mathbf{n}$ is the number of control intervals from $t$ when a near user is scheduled. Thus at control intervals $\{t, t-1, \ldots 1\} - \{t_{\mathbf{n}}, t_{\mathbf{n}-1}, \ldots t_1\}$, a far user is scheduled in cell 1. Define $\mathbf{t}_k := \{t_k, t_{k-1}, \ldots t_1\}$ with $k \leq \mathbf{n}$. Note that, in the sporadic time axis $\mathbf{t_n}$ ($\{t, t-1, \ldots 1\} - \mathbf{t_n}$), base station 2, in order to maintain cell breathing, schedules far (near) users.

Let $\mathfrak{A}^f$ and $\mathfrak{A}^n$ denote the scheduling policies adopted by base station 2 in the sporadic time axes corresponding, respectively, to near and far schedule decisions in cell 1. Let $\{\hat{\mathfrak{A}}, \mathfrak{A}^f, \mathfrak{A}^n\}$ denote the two-cell scheduling policy with $\hat{\mathfrak{A}}$ indicating the use of greedy policy in cell 1. We now introduce the following lemma.

**Lemma 1.** *Under the asymmetric cooperation assumption, if, for any fixed $t$, for every realization of $\{\mathbf{n}, \mathbf{t_n}\}$, the scheduling policy $\{\mathfrak{A}^f, \mathfrak{A}^n\}$ is throughput maximizing in cell 2, then $\{\hat{\mathfrak{A}}, \mathfrak{A}^f, \mathfrak{A}^n\}$ is sum throughput optimal.*

*Proof.* The lemma is not obvious due to a possible influence of the policies $\mathfrak{A}^f$ and $\mathfrak{A}^n$ on the sporadic time axis $\mathbf{t_n}$, potentially invalidating the realization based argument. Under the asymmetric cooperation assumption, since base station 1 makes near/far schedule decisions without consulting base station 2 and since the channel states evolve independently at the underlying physical layer [4], $\{\mathbf{n}, \mathbf{t_n}\}$ is independent of the schedule decisions and observations made in cell 2 within the sporadic time axes and hence is independent of $\mathfrak{A}^f$ and $\mathfrak{A}^n$. This decoupling along with the fact that the greedy policy is optimal in cell 1 establishes the lemma. □

We now proceed to show that the greedy policy is optimal within a realization of the sporadic time axes. Note that the greedy policy was shown to be optimal on a non-sporadic time axis in [9, 10]. However, in the current case, since the belief values evolve across non-uniform time steps, we need a rigorous optimality proof in the changed setting.

Fix a realization $\{\mathbf{n}, \mathbf{t_n}\}$ throughout the following analysis. The net expected reward accrued by base station 2 from $t_{k \leq \mathbf{n}}$ on the time axis $\mathbf{t_n}$ is given as follows.

$$V_{t_k}(\pi_{t_k}, \{a_{t_k}, \{\mathfrak{A}^f_{t_l}\}_{k>l>0}\}) \;=\; \pi_{t_k}(a_{t_k}) + \mathrm{E}\left[V_{t_{k-1}}(\pi_{t_{k-1}}, \{\mathfrak{A}^f_{t_l}\}_{k-1>l>0} | \pi_{t_k}, a_{t_k})\right] \quad (8)$$

where $\pi_{t_k}$ is the vector of the channel states of the far users in cell 2 in control interval $t_k$, $a_{t_k}$ is the far user scheduled in control interval $t_k$ and $\{\mathfrak{A}^f_{t_l}\}_{k>l>0}$ is the scheduling policy from the next instant till the horizon on $\mathbf{t_n}$. We now establish the structure of the greedy policy on the sporadic time axis. In any control interval $t_k$, $k < \mathbf{n}$, the belief values of the users are given as follows.

$$\pi_{t_k}(i) = \begin{cases} p, & \text{if } i = a_{t_{k+1}}, F_{t_{k+1}} = 1 \\ r, & \text{if } i = a_{t_{k+1}}, F_{t_{k+1}} = 0 \\ T^{(t_{k+1}-t_k)}(\pi_{t_{k+1}}(i)), & \text{if } i \neq a_{t_{k+1}}. \end{cases} \quad (9)$$

where $F_t$ is the ARQ feedback received by base station 2 from the scheduled user in control interval $t$ with 1(0) corresponding to a ACK(NACK). Note that for $0 \leq x \leq 1$, $T(x) = x(p-r) + r$. Thus $T(x) \in [r, p]$. Since $T^l(x) = T(T^{l-1}(x))$, by induction,

[4]Note that the inter-cell, intra-cell *base station to user* links are assumed to be statistically independent.

$T^l(x) \in [r,p]$, when $l > 0$. Also $T(x_1) \geq T(x_2)$ if $x_1 \geq x_2$. Thus, by induction, $T^l(x_1) \geq T^l(x_2)$ if $x_1 \geq x_2$. We now introduce the schedule order vector $O_{t_k}$ as the ordered arrangement of the index of the users in decreasing order of $\pi_{t_k}(i)$, i.e.,

$$\begin{aligned} O_{t_k}(1) &= \arg\max_i \pi_k(i) \\ &\vdots \\ O_{t_k}(N_f) &= \arg\min_i \pi_k(i). \end{aligned}$$

where $N_f$ is the number of far users in cell 2. From the preceding discussion on the structure of $T(x)$ and the evolution of the belief values, the schedule order vector evolves as below:

$$O_{t_{k-1}} = \begin{cases} [a_{t_k}\ \{O_{t_k} - a_{t_k}\}], & \text{if } f_{t_k} = 1 \\ [\{O_{t_k} - a_{t_k}\}\ a_{t_k}], & \text{if } f_{t_k} = 0, \end{cases} \tag{10}$$

The greedy policy which aims to maximize the immediate reward (belief value), picks the user on the top of the schedule order vector and thus has a round-robin structure, with user switch triggered by a NACK, on the sporadic time axis, as observed on the non-sporadic axis [9, 10]. We now proceed to show the optimality of the greedy policy on $\mathbf{t_n}$ by first deriving a sufficient condition for optimality, similar to our analysis in [all].

Consider a control interval $t_m, m \leq \mathbf{n}$ with belief vector $\pi_{t_m}$ and action $a_{t_m}$. Let the users be indexed in the order of their belief values in control interval $t_m$, i.e, $O_{t_m} = [1 \ldots N_f]$. Assuming $\{\mathfrak{A}_{t_k}\}_{k \leq m-1} = \{\widehat{\mathfrak{A}}_{t_k}\}_{k \leq m-1}$. Let $S_{t_k}$, the state vector, denote the 1/0 channel states of the users at $t_k$. We write the net expected reward as follows

$$\begin{aligned} &V_{t_m}(\pi_{t_m}, \{a_{t_m}, \{\widehat{\mathfrak{A}}_{t_k}\}_{k \leq m-1}\}) \\ &= \pi_{t_m}(a_{t_m}) + \sum_{S_{t_m}} P_{S_{t_m}|\pi_{t_m}}(S_{t_m}|\pi_{t_m}) \hat{V}_{t_{m-1}}(S_{t_m}, O_{t_{m-1}}), \end{aligned}$$

where $\hat{V}_{t_{m-1}}$ is the expected future reward conditioned on the state vector in the previous control interval on the sporadic time axis, i.e., $t_m$. The *hat* on this quantity emphasizes the use of the greedy policy in all $t_k, k \leq m-1$. $P_{S_{t_m}|\pi_{t_m}}(S_{t_m}|\pi_{t_m})$ is the conditional probability of the current state vector $S_{t_m}$ given the belief vector $\pi_{t_m}$. Note that the schedule order vector $O_{t_{m-1}}$ is only a function of $O_{t_m}$ and the state $S_{t_m}(a_{t_m})$, thus maintaining consistency with the amount of information available for scheduling decision in the actual problem setup. We now proceed to compare the net expected reward when $a_{t_m} = n$ and $a_{t_m} = n+1$ where $n \in \{1 \ldots N_f - 1\}$. Let $Y$ and $X$ be random binary vectors of lengths

$n-1$ and $N_f - n - 1$ (empty when the length is non-positive) respectively. Then,

$$
\begin{aligned}
& V_{t_m}(\pi_{t_m}, \{a_{t_m} = n, \{\widehat{\mathfrak{A}}_{t_k}\}_{k \leq m-1}\}) \\
& \quad = \pi_{t_m}(n) + \sum_{Y,X} P_{S_{t_m}|\pi_{t_m}}([Y\ 0\ 0\ X]|\pi_{t_m}) \times \hat{V}_{t_{m-1}}\big([Y\ 0\ 0\ X], [\{O_{t_m} - n\}\ n]\big) \\
& \qquad + \sum_{Y,X} P_{S_{t_m}|\pi_{t_m}}([Y\ 0\ 1\ X]|\pi_{t_m}) \times \hat{V}_{t_{m-1}}\big([Y\ 0\ 1\ X], [\{O_{t_m} - n\}\ n]\big) \\
& \qquad + \sum_{Y,X} P_{S_{t_m}|\pi_{t_m}}([Y\ 1\ 0\ X]|\pi_{t_m}) \times \hat{V}_{t_{m-1}}\big([Y\ 1\ 0\ X], [n\ \{O_{t_m} - n\}]\big) \\
& \qquad + \sum_{Y,X} P_{S_{t_m}|\pi_{t_m}}([Y\ 1\ 1\ X]|\pi_{t_m}) \times \hat{V}_{t_{m-1}}\big([Y\ 1\ 1\ X], [n\ \{O_{t_m} - n\}]\big),
\end{aligned} \tag{11}
$$

Since the Markov channel statistics are identical across the users, we have the following symmetry property: for any $k \geq 1$,

$$
\begin{aligned}
& \hat{V}_{t_k}(S_{t_{k+1}}, O_{t_k}) = \hat{V}_{t_k}(\tilde{S}_{t_{k+1}}, \tilde{O}_{t_k}) \\
\text{if} \quad & S_{t_{k+1}}(O_{t_k}(i)) = \tilde{S}_{t_{k+1}}(\tilde{O}_{t_k}(i)) \ \ \forall\ i \in \{1 \ldots N_f\}.
\end{aligned} \tag{12}
$$

Expanding $V_{t_m}(\pi_{t_m}, \{a_{t_m} = n+1, \{\widehat{\mathfrak{A}}_{t_k}\}_{k \leq m-1}\})$ along the lines of (11), and using the symmetry property, with further mathematical simplification, we can evaluate the difference in the net expected reward as follows,

$$
\begin{aligned}
& V_{t_m}(\pi_{t_m}, \{a_{t_m} = n, \{\widehat{\mathfrak{A}}_{t_k}\}_{k \leq m-1}\}) - V_{t_m}(\pi_{t_m}, \{a_{t_m} = n+1, \{\widehat{\mathfrak{A}}_{t_k}\}_{k \leq m-1}\}) \\
& \quad = \Big(\pi_{t_m}(n) - \pi_{t_m}(n+1)\Big)\Big(1 - \sum_{Y,X} \Big[\big[\hat{V}_{t_{m-1}}\big([Y\ 1\ X\ 0], [1 \ldots N_f]\big) \\
& \qquad - \hat{V}_{t_{m-1}}\big([1\ Y\ 0\ X], [1 \ldots N_f]\big)\big] \times P_{S_{t_m}|\pi_{t_m}}\big([S_{t_m}(1) \ldots S_{t_m}(n-1)] = Y|\pi_{t_m}\big) \times \\
& \qquad P_{S_{t_m}|\pi_{t_m}}\big([S_{t_m}(n+2) \ldots S_{t_m}(N_f)] = X|\pi_{t_m}\big)\Big]\Big).
\end{aligned} \tag{13}
$$

**Lemma 2.** *Greedy policy maximizes the throughput on the sporadic time axis* $\mathbf{t_n}$ *if the following (sufficient) condition holds.*

$$
\hat{V}_{t_{m-1}}\big([Y\ 1\ X\ 0], [1 \ldots N_f]\big) - \hat{V}_{t_{m-1}}\big([1\ Y\ 0\ X], [1 \ldots N_f]\big) \ \leq \ 1, \tag{14}
$$

$\forall\ \mathbf{n} \geq m > 1$, $n \in \{1 \ldots N_f - 1\}$, $Y$, $X$ *being random binary vectors of length* $n-1$ *and* $N_f - n - 1$ *and* $\hat{V}_{t_{m-1}}$ *is the reward accrued under the greedy policy, i.e., when* $\{\mathfrak{A}_{t_k}\}_{k \leq m-1} = \{\widehat{\mathfrak{A}}_{t_k}\}_{k \leq m-1}$.

*Proof.* Let condition (16) be true. Let $m > 1$ be fixed. Since, by assumption, $\pi_{t_m}(n) \geq \pi_{t_m}(n+1)\ \forall n \in \{1 \ldots N_f - 1\}$, we have from (13),

$$
V_{t_m}(\pi_{t_m}, \{a_{t_m} = n, \{\widehat{\mathfrak{A}}_{t_k}\}_{k \leq m-1}\}) \ \geq \ V_{t_m}(\pi_{t_m}, \{a_{t_m} = n+1, \{\widehat{\mathfrak{A}}_{t_k}\}_{k \leq m-1}\}).
$$

Therefore,

$$\begin{aligned} &V_{t_m}(\pi_{t_m}, \{a_{t_m} = \arg\max_i \pi_{t_m}(i) = 1, \{\widehat{\mathfrak{A}}_{t_k}\}_{k \leq m-1}\}) \\ &\geq \; V_{t_m}(\pi_{t_m}, \{a_{t_m} \in \{2 \ldots N\}, \{\widehat{\mathfrak{A}}_{t_k}\}_{k \leq m-1}\}). \end{aligned}$$

We now have the following statement:

If $\forall \pi_{t_{m-1}} \in [0,1]^{N_f}$,

$$\{\widehat{\mathfrak{A}}_{t_k}\}_{k \leq m-1} = \arg \max_{\{\mathfrak{A}_{t_k}\}_{k \leq m-1}} V_{t_{m-1}}(\pi_{t_{m-1}}, \{\mathfrak{A}_{t_k}\}_{k \leq m-1}),$$

then $\forall \pi_{t_m} \in [0,1]^{N_f}$,

$$\{\widehat{\mathfrak{A}}_{t_k}\}_{k \leq m} = \arg \max_{\{\mathfrak{A}_{t_k}\}_{k \leq m}} V_{t_m}(\pi_{t_m}, \{\mathfrak{A}_{t_k}\}_{k \leq m}). \tag{15}$$

Since $\widehat{\mathfrak{A}}_{t_1} = \arg\max_{\mathfrak{A}_{t_1}} V_{t_1}(\pi_{t_1}, \mathfrak{A}_{t_1}), \forall \pi_{t_1} \in [0,1]^{N_f}$, using (15), by induction, we have

$$\begin{aligned} \{\widehat{\mathfrak{A}}_{t_k}\}_{k \leq m} \; &= \; \arg \max_{\{\mathfrak{A}_{t_k}\}_{k \leq m}} V_{t_m}(\pi_{t_m}, \{\mathfrak{A}_{t_k}\}_{k \leq m}) \\ &\qquad \forall \mathbf{n} \geq m \geq 1, \pi_{t_m} \in [0,1]^{N_f}. \end{aligned}$$

The lemma thus follows. □

We now formally introduce the optimal multiuser scheduling policy in the two-cell system with asymmetric cooperation.

**Proposition 3.** *The policy* $\{\hat{\mathfrak{A}}, \hat{\mathfrak{A}}^f, \hat{\mathfrak{A}}^n\}$ *maximizes the sum throughput of the two-cell system.*

*Proof.* We begin by establishing that the sufficient condition in fact holds, using the round robin structure of the greedy policy on the sporadic time axis. Consider a realization of the channel states of the $N_f$ users on the time axis $\mathbf{t}_{m-1}$, $m \leq n$. Denote it by $\{R, i, j\}$, where $i, j$ indicate the channel state of users $n+1$ and $N_f$, respectively, at time $t_{m-1}$ with $R$ indicating the rest of the channel state realization. We can rewrite the second quantity of the sufficient condition as follows.

$$\hat{V}_{t_{m-1}}\big([1 \; Y \; 0 \; X], [1 \ldots N_f]\big) \;\; = \;\; \hat{V}_{t_{m-1}}\big([Y \; 0 \; X \; 1], [N_f, 1 \ldots N_f - 1]\big)$$

Define $V_a(\{R, i, j\})$ as the reward accrued from time $t_{m-1}$ on the sporadic axis when the channel states have a realization $\{R, i, j\}$ and the greedy policy is implemented in the order $[1 \ldots N_f]$ from control interval $t_{m-1}$. Let $V_b(\{R, i, j\})$ be similarly defined with the order given by $[N_f, 1 \ldots N_f - 1]$. Let $\mathcal{P}(\{i, j\} | \{k, l\}) = P(S_{t_{m-1}}(n+1) = i, S_{t_{m-1}}(N_f) =$ $j|S_{t_m}(n+1)=k, S_{t_m}(N_f)=l)$. The sufficient condition can now be rewritten as below.

$$\begin{aligned}
&\hat{V}_{t_{m-1}}\big([Y\ 1\ X\ 0],[1\dots N_f]\big)-\hat{V}_{t_{m-1}}\big([1\ Y\ 0\ X],[1\dots N_f]\big)\\
&= \sum_R P(R|S_{t_m}(1)\dots S_{t_m}(n)=Y, S_{t_m}(n+2)\dots S_{t_m}(N_f)=X)\times\\
&= \mathcal{P}(\{1,0\}|\{1,0\})V_a(\{R,1,0\})-\mathcal{P}(\{0,1\}|\{0,1\})V_b(\{R,0,1\})\\
&\quad +\mathcal{P}(\{0,1\}|\{1,0\})V_a(\{R,1,0\})-\mathcal{P}(\{1,0\}|\{0,1\})V_b(\{R,0,1\})\\
&\quad +\mathcal{P}(\{0,0\}|\{1,0\})V_a(\{R,1,0\})-\mathcal{P}(\{0,0\}|\{0,1\})V_b(\{R,0,1\})\\
&\quad +\mathcal{P}(\{1,1\}|\{1,0\})V_a(\{R,1,0\})-\mathcal{P}(\{1,1\}|\{0,1\})V_b(\{R,0,1\})\\
&= p(1-r)(V_a(\{R,1,0\})-V_b(\{R,0,1\}))+(1-p)(r)(V_a(\{R,0,1\})-V_b(\{R,1,0\}))\\
&\quad +(1-p)(1-r)(V_a(\{R,0,0\})-V_b(\{R,0,0\}))+pr(V_a(\{R,1,1\})-V_b(\{R,1,1\}))
\end{aligned} \tag{16}$$

It has been shown in [10] that when greedy policy is implemented in orders $[1\dots N_f]$ and $[N_f, 1\dots N_f-1]$, the difference in reward accrued, for any fixed realization, is upper bounded by 1. The sample path argument used in the proof works for the non-sporadic axis as well. Thus $V_a(\{R,i,j\})-V_b(\{R,i,j\})\le 1$ for any $\{R,i,j\}$. Notice that since the realization is fixed and since $V_a(\{R,i,j\})$ schedules user 1 first, the value of $j$ does not affect $V_a$. Thus $V_a(\{R,i,1\})=V_a(\{R,i,0\})$. Similarly $V_b(\{R,1,j\})=V_a(\{R,0,j\})$. Using these observations in (16), we show the sufficient condition holds. The proposition is thus established from Lemma 1 and Lemma 2. □

## 3.2 Under Symmetric Cooperation With Constraints on the Breathing Pattern

Assume the cells cooperate mutually in maintaining the cell breathing protocol. Assume there is a constraint on the breathing pattern, i.e., cell 1 *must* breathe-out (schedule far users) and cell 2 must breathe-in in a predetermined exhaustive sequence of control intervals $\mathbf{t}$. We have the following observation.

**Proposition 4.** *Under a fixed* $\mathbf{t}$*, the optimal scheduling policies are decoupled across the cells. The optimal policies within each group in each cell is a greedy policy.*

*Proof.* We see this readily from the following two observations:

- Decoupling: with $\mathbf{t}$ fixed, within $\mathbf{t}$, the schedule decision of a cell does not affect the schedule decisions of the neighboring cells, since there is no burden of maintaining the cell breathing protocol, any more. The same argument holds for the complementary time axis.
- We have earlier shown that, for a fixed realization of the sporadic time axis, if a cell has to schedule an user from only one group, then greedy is optimal within that axis.

□

We have a straightforward corollary to this result.

**Corollary 5.** *When there are multiple orthogonal carriers available to the cells, with users assigned to the carriers such that no single carrier serves a near-near or far-far pair across the cells, then it is optimal to greedily schedule users within every carrier within each cell.*

Note that the decoupling argument of proposition is used here. Availability of multiple carriers is usually the case in narrowband systems.

# 4 Restless Multiarmed Bandit Processes

A direct analysis of the ARQ based scheduling problem with symmetric cooperation and no constraints on the breathing pattern appears very difficult due to the complex relationship between the schedule decisions across space and time. We therefore establish a connection between the scheduling problem and the restless multiarmed bandit processes (RMAB) and make use of the established theory behind RMAB in our analysis. We proceed with a survey on the RMAB theory.

Multi-armed bandit problems [23] are defined as a family of sequential dynamic resource allocation problems in the presence of several competing, *independently evolving* projects. They are characterized by a fundamental tradeoff between decisions guaranteeing high immediate rewards versus those that sacrifice immediate rewards for better future rewards. Several technological and scientific disciplines such as sensor management, manufacturing systems, economics, queueing and communication networks ([23]) encounter resource allocation problems that can be modeled as MAB processes. In the classic MAB processes, in each control interval, a single project has to be allocated the available system resources. The state of the project thus scheduled evolves from the current time slot to the next time slot. Whereas, those not scheduled remain frozen. Gittins and Jones ([24]) studied these processes and showed that the optimal solution is of the index type: i.e., for each bandit process (project or an arm of the MAB), an index that is a function of the state of the project is computed and the project with the highest index is scheduled. The index was called by the authors as the Dynamic Allocation Index, but is now justly known as *Gittins index*. Note that the optimal scheduling policy, that originally required the solution of a $N$-armed bandit process ($N$ being the number of projects), is now reduced to determining the Gittins index for $N$ single armed bandit processes, thus exponentially reducing the problem complexity. This complexity reduction is one of the main reasons behind the immense interest in index policies for MAB processes and its variants. We will discuss this next.

Whittle [25] generalized the MAB processes as follows: In each control interval, exactly $M \geq 1$ projects are scheduled. The states of the rest $N-M$ projects are not frozen like in MAB, but evolve in time. They also contribute rewards ($W$) known as *passivity* rewards. These processes are called the *Restless* multiarmed bandit processes (RMAB), the term restless being indicative of the state evolution of even the projects that are not scheduled at the moment. Considering a single project, Whittle defines the $W$-*subsidy* policy as follows: In each control interval, schedule the project if the sum of the immediate reward and the future reward corresponding to an active decision outweighs the sum of the reward for passivity ($W$) and the corresponding future reward. For a state $\pi$ of the project, the value of $W$ corresponding to equal net rewards for active and passive decisions is defined as the index $I(\pi)$. The notion of *Indexability* is now defined by Whittle as below:

*Let $D(W)$ be the set of states for which a project would be made passive under a $W$-subsidy policy. The project is indexable if $D(W)$ increases monotonically from $\phi$ to $\mathcal{S}$ as $W$ increases from $-\infty$ to $\infty$*

where $\phi$ is the empty set and $\mathcal{S}$ is the universal set of the states of the project. The notion of indexability gives a consistent ordering of states with respect to the indices, i.e., if $I(\pi_1) > I(\pi_2)$ and if it is optimal to activate the project when in state $\pi_2$, then it is optimal to activate the project when it is in state $\pi_1$. Returning to the RMAB scheduling problem, Whittle proposes the *index* scheduling policy: In each control interval, activate the $M$ projects that have the greatest indices. Note that the natural ordering of states based on indices (under indexability) gives credibility to the index policy. Whittle shows that when the strict constraint on the number of projects per interval ($M$) is relaxed to an average constraint, the index scheduling policy is optimal. He also shows that when the *restlessness* aspect is removed from the RMAB and when $W = 0$, the index reduces to the Gittins index. He conjectured that, with $\frac{M}{N}$ fixed, as $M, N \to \infty$, the index policy is optimal. This was later proved to be true in [26] except for very special cases of RMAB processes.

Indexability is a very strict requirement ([25]) that is hard to check. There have been several works [26, 27, 28, 29, 30, 31] on indexability and index policies for various RMAB processes. In [26], for a special class of RMAB, the authors show that, if the RMAB is indexable, under certain technical conditions, the index policy is optimal. In [27], the authors provide a sufficient condition for the indexability of a single restless bandit. The authors in [29] investigate indexability under a set of conditions called Partial Conservation Laws (PCL). They identify a class of RMAB processes that satisfy the PCL and are indexable in the sense of Whittle. They also show that, under PCL, if the rewards belong to a certain "admissible region" then a priority index based allocation policy is optimal. In [31], the authors consider a RMAB process with improving/deteriorating jobs and establish indexability for the processes. They demonstrate, via numerical analysis, the strong performance of the index policy and obtain performance guarantees for the index policy. Thus we see that the notion of indexability offers a promising starting point towards the (otherwise intractable) analysis of optimal RMAB scheduling.

Returning to the scheduling problem addressed in this work, assume that the near users in cell 1 are permanently paired (one to one) with far users in cell 2 and vice versa (requires $N = F$). Thus if a user is scheduled in a cell, under cell breathing its pair must be scheduled in the adjacent cell. We can now visualize each pair as a restless bandit with one and only one pair scheduled in any control interval. Thus the ARQ based scheduling problem is now a RMAB scheduling problem. When the permanent pairing condition is removed, we have a set of $2NF$ projects, considering all possible legitimate pairing across cells. These projects do not evolve independently and hence do not constitute a RMAB process[5]. Thus we have a more complex variant of the RMAB processes. We call this the RMAB-v processes. To the best of our knowledge, there is no analysis of scheduling in RMAB-v processes.

Note, from previous discussion, that the index policies are very attractive from an implementation point of view. From an optimality point of view, the attractiveness of the index policies can be attributed to the natural ordering of states (and hence projects) based on indices, as guaranteed by indexability. We are curious to see whether this advantage carries over to the RMAB-v processes at hand. As a first step in this direction, we perform an indexability analysis of the RMAB-v and obtain partial results on the structure of the index policy in the following sections.

# 5 Indexability of the RMAB-v

Following the approach of Whittle in [25], we consider the indexability of a single project made by a near-far user pair. We begin with the following definition: A function $f(x_1, x_2, \dots x_n)$ is component-wise piecewise linear and component-wise convex in $(x_1, x_2, \dots x_n)$ if, by fixing arbitrary values along any $n-1$ dimensions, $f(.)$ is piecewise linear and convex in the remaining dimension.

**Lemma 6.** *The reward function, $V_t(W, \pi_1, \pi_2)$, is component-wise piecewise linear and component-wise convex in $(W, \pi_1, \pi_2)$, for any $t \geq 1$.*

*Proof.* Let $\mathcal{F}$ denote the family of functions defined over $(W, \pi_1, \pi_2) \in \mathbb{R}^+ \times [0, 1]^2$ that are component-wise piecewise linear and component-wise convex in $(W, \pi_1, \pi_2)$. Let $L(x)$ be an arbitrarily defined affine function in $x \in [0, 1]$. The reward function at control interval $t$, when the state of the system is $(L(\pi_1), L(\pi_2))$, is given by

$$\begin{aligned} V_t(W, L(\pi_1), L(\pi_2)) \;=\; \max\Big(& W + V_{t-1}(W, T(L(\pi_1)), T(L(\pi_2))), L(\pi_1) + L_1(\pi_2) \\ & + L(\pi_1)L(\pi_2)V_{t-1}(W, p, p) + L(\pi_1)(1 - L(\pi_2))V_{t-1}(W, p, r) \\ & + (1 - L(\pi_1))L(\pi_2)V_{t-1}(W, r, p) \\ & + (1 - L(\pi_1))(1 - L(\pi_2))V_{t-1}(W, r, r)\Big) \end{aligned} \tag{17}$$

Let (A) denote the following condition: $V_{t-1}(W, L(\pi_1), L(\pi_2)) \in \mathcal{F}$ for any arbitrarily defined affine function $L$.

Note that $T(L(x))$ is affine in $x$. Hence, under (A), both the arguments to the max operator are component-wise piecewise linear and component-wise convex in $(W, \pi_1, \pi_2)$. Since the max operator, across each dimension, effectively traces the top envelope of a set of piecewise linear, convex functions, the piecewise linearity and convexity is preserved across each dimension. Thus $V_t(W, L(\pi_1), L(\pi_2)) \in \mathcal{F}$ under condition (A).

We now proceed to establish the induction basis for condition (A). Since, by definition, $V_1(W, L(\pi_1), L(\pi_2)) = \max(W, L(\pi_1) + L(\pi_2))$, we readily see that $V_1(W, L(\pi_1), L(\pi_2)) \in \mathcal{F}$. Thus, by induction, from the preceding statements, $V_t(W, L(\pi_1), L(\pi_2)) \in \mathcal{F}$, $\forall t \geq 1$.

With $L(x) = x$, the lemma is established. ∎

[5]The projects must evolve independently in RMAB processes, by definition.

We now establish a relation between the expected future rewards accrued after active (schedule) and passive (idle) decisions. Let $V_t^A(W,\pi_1,\pi_2)$ and $V_t^P(W,\pi_1,\pi_2)$, for $t \geq 2$, denote the expected future rewards (accrued from control interval $t-1$) corresponding to active and passive decisions, respectively, at control interval $t$, with $(\pi_1,\pi_2)$ being the state of the system at $t$. By definition, we have

$$\begin{aligned} V_t^A(W,\pi_1,\pi_2) &= \pi_1\pi_2 V_{t-1}(W,p,p) + \pi_1(1-\pi_2)V_{t-1}(W,p,r) \\ &\quad +(1-\pi_1)\pi_2 V_{t-1}(W,r,p) + (1-\pi_1)(1-\pi_2)V_{t-1}(W,r,r) \\ V_t^P(W,\pi_1,\pi_2) &= V_{t-1}(W,T(\pi_1),T(\pi_2)). \end{aligned} \tag{18}$$

**Lemma 7.** *The expected future reward accrued after an active decision is at least as high as the expected future reward accrued after a passive decision, i.e.,* $V_t^A(W,\pi_1,\pi_2) \geq V_t^P(W,\pi_1,\pi_2)$, $\forall t \geq 2, W \geq 0, (\pi_1,\pi_2) \in [0,1]^2$.

*Proof.* We begin by rewriting $V_t^A(W,\pi_1,\pi_2)$:

$$\begin{aligned} V_t^A(W,\pi_1,\pi_2) &= \pi_2(\pi_1 V_{t-1}(W,p,p) + (1-\pi_1)V_{t-1}(W,r,p)) \\ &\quad +(1-\pi_2)(\pi_1 V_{t-1}(W,p,r) + (1-\pi_1)V_{t-1}(W,r,r)) \\ &\geq \pi_2 V_{t-1}(W,\pi_1 p + (1-\pi_1)r, p) + (1-\pi_2)V_{t-1}(W,\pi_1 p + (1-\pi_1)r, r) \\ &\geq V_{t-1}(W,\pi_1 p + (1-\pi_1)r, \pi_2 p + (1-\pi_2)r) \\ &= V_{t-1}(W,T(\pi_1),T(\pi_2)) \\ &= V_t^P(W,\pi_1,\pi_2) \end{aligned} \tag{19}$$

where the second and third inequalities use the component-wise convexity of $V_{t-1}$ along the $\pi_1$ and $\pi_2$ dimensions, respectively (Lemma 6). □

**Lemma 8.** *When $W \leq 2r$ or $W \geq 2p$, the expected future rewards accrued after active and passive decisions are equal, i.e.,* $V_t^A(W,\pi_1,\pi_2) = V_t^P(W,\pi_1,\pi_2)$, $\forall t \geq 2, (\pi_1,\pi_2) \in [0,1]^2$, $W \notin (2r,2p)$.

*Proof.* We consider the following two cases.

When $W \leq 2r$:
Recall from (18),

$$\begin{aligned} V_t^A(W,\pi_1,\pi_2) &= \pi_1\pi_2 V_{t-1}(W,p,p) + \pi_1(1-\pi_2)V_{t-1}(W,p,r) \\ &\quad +(1-\pi_1)\pi_2 V_{t-1}(W,r,p) + (1-\pi_1)(1-\pi_2)V_{t-1}(W,r,r) \end{aligned} \tag{20}$$

Note that $V_{t-1}(W,r,r) = \max(W + V_{t-1}^P(W,r,r), 2r + V_{t-1}^A(W,r,r))$. Since, from Lemma 7, $V_{t-1}^A(W,\pi_1,\pi_2) \geq V_{t-1}^P(W,\pi_1,\pi_2)$, $\forall W > 0, (\pi_1,\pi_2) \in [0,1]^2$, we have $V_{t-1}(W,r,r) = 2r + V_{t-1}^A(W,r,r)$ when $W \leq 2r$. Using the same argument for the other elements in (20), we have

$$\begin{aligned} V_t^A(W,\pi_1,\pi_2) &= \pi_1\pi_2(2p + V_{t-1}^A(W,p,p)) + \pi_1(1-\pi_2)(p + r + V_{t-1}^A(W,p,r)) \\ &\quad +(1-\pi_1)\pi_2(r + p + V_{t-1}^A(W,r,p)) \\ &\quad +(1-\pi_1)(1-\pi_2)(2r + V_{t-1}^A(W,r,r)) \end{aligned} \tag{21}$$

Recall from (18),

$$\begin{aligned} V_t^P(W,\pi_1,\pi_2) &= V_{t-1}(W,T(\pi_1),T(\pi_2)) \\ &= \max\Big(W + V_{t-1}^P\big(W,T(\pi_1),T(\pi_2)\big), \\ &\quad T(\pi_1)+T(\pi_2)+V_{t-1}^A\big(W,T(\pi_1),T(\pi_2)\big)\Big) \end{aligned} \tag{22}$$

Note that $2r \leq T(\pi_1) + T(\pi_2) = \pi_1 p + (1-\pi_1)r + \pi_2 p + (1-\pi_2)r \leq 2p$. Thus, with $V_{t-1}^A\big(W,T(\pi_1),T(\pi_2)\big) \geq V_{t-1}^P\big(W,T(\pi_1),T(\pi_2)\big)$, when $W \leq 2r$,

$$\begin{aligned} V_t^P(W,\pi_1,\pi_2) &= T(\pi_1)+T(\pi_2)+V_{t-1}^A(W,T(\pi_1),T(\pi_2)) \\ &= \pi_1(p-r)+r+\pi_2(p-r)+r \\ &\quad +(\pi_1(p-r)+r)(\pi_2(p-r)+r)V_{t-2}(W,p,p) \\ &\quad +(\pi_1(p-r)+r)(1-(\pi_2(p-r)+r))V_{t-2}(W,p,r) \\ &\quad +(1-(\pi_1(p-r)+r))(\pi_2(p-r)+r)V_{t-2}(W,r,p) \\ &\quad +(1-(\pi_1(p-r)+r))(1-(\pi_2(p-r)+r))V_{t-2}(W,r,r) \end{aligned} \tag{23}$$

Expanding $V_{t-1}^A(W,\{r,p\},\{r,p\})$ in (21) using (18) along with the fact[6] that $V_t(W,p,r) = V_t(W,r,p)$, we equate (21) and (23), thus establishing the first part of the lemma.

When $W \geq 2p$:

Let $\pi_1,\pi_2$ be the belief value in any control interval $t>1$. Then, in any control interval $m<t$, $\pi_1|_{m<t} \in \{r,p,T^j(\pi_1),T^k(r),T^l(p)\}$ and $\pi_2|_{m<t} \in \{r,p,T^j(\pi_2),T^k(r),T^l(p)\}$ with $j \leq t-1$ and $k,l \leq t-2$. Note that the belief values $\pi_1,\pi_2|_{m<t} \in \{r,p\}$ if the schedule decision in the previous control interval was active. While the belief values take the other forms when there is a continuous stretch of passive decisions immediately preceding the control interval in question. By the definition of $T^k$, we can see that $r \leq T^j(\pi) \leq p$, $\forall \pi \in [0,1]$ and $j \geq 1$. Thus we see that, $\forall t < t_0, 2r \leq \pi_1 + \pi_2|_{m<t} \leq 2p$. With this observation consider, from (18),

$$\begin{aligned} V_t^A(W,\pi_1,\pi_2) &= \pi_1\pi_2 V_{t-1}(W,p,p) + \pi_1(1-\pi_2)V_{t-1}(W,p,r) \\ &\quad +(1-\pi_1)\pi_2 V_{t-1}(W,r,p) + (1-\pi_1)(1-\pi_2)V_{t-1}(W,r,r) \end{aligned} \tag{24}$$

Consider $V_{t-1}(W,p,p)$ from the preceding equation. From the preceding discussion, $\forall m < t-1$, $2r \leq \pi_1 + \pi_2|_{m<t-1} \leq 2p$. Also note that $\pi_1 + \pi_2|_{t-1} = 2p$. Thus for any sequence of schedule choices and ARQ feedback, the immediate reward in any control interval $m \leq t-1$ is upper bounded by $2p$. Thus when $W \geq 2p$, it is optimal to not schedule in all control intervals from $t-1$. Thus $V_{t-1}(W \geq 2p,p,p) = (t-1)W$. Using a similar argument we have $V_{t-1}(W \geq 2p,p,r) = V_{t-1}(W \geq 2p,r,p) = V_{t-1}(W \geq 2p,r,r) = (t-1)W$. Therefore, from (24),

$$V_t^A(W,\pi_1,\pi_2) = (t-1)W, \quad \forall (\pi_1,\pi_2) \in [0,1]^2, W \geq 2p \tag{25}$$

[6]Thanks to the channel, reward structure symmetry across the users

From (18), the future reward after passive decision is given by

$$V_t^P(W, \pi_1, \pi_2) \;=\; V_{t-1}(W, T(\pi_1), T(\pi_2)) \quad (26)$$

Since $2r \leq T(\pi_1) + T(\pi_2) \leq 2p$, $\forall (\pi_1, \pi_2) \in [0, 1]^2$, using an argument similar to above, we have

$$V_t^P(W, \pi_1, \pi_2) \;=\; (t-1)W, \;\; \forall (\pi_1, \pi_2) \in [0, 1]^2, W \geq 2p \quad (27)$$

This completes the proof. □

**Lemma 9.** *If for any $t > 1, (\pi_1, \pi_2) \in [0, 1]^2$, $W^*$ is such that $W + V_t^P(W, \pi_1, \pi_2) = \pi_1 + \pi_2 + V_t^A(W, \pi_1, \pi_2)|_{W=W^*}$, then the project is indexable (in the sense of Whittle) at time $t$ iff $W^*$ is unique. The index is given by $I(t, \pi_1, \pi_2) = W^*$.*

*Proof.* We first consider the sufficiency part. For a fixed $t, (\pi_1, \pi_2) \in [0, 1]^2$, at $W = W^*$, $W - (\pi_1 + \pi_2) = V_t^A(W, \pi_1, \pi_2) - V_t^P(W, \pi_1, \pi_2)$. Note that, from Lemma 7, $V_t^A(W, \pi_1, \pi_2) - V_t^P(W, \pi_1, \pi_2) \geq 0 \; \forall W > 0$. Also, $W - (\pi_1 + \pi_2)$ is an increasing function in $W$ with negative values $\forall \; W < (\pi_1 + \pi_2)$. Using these, along with the uniqueness of $W^*$, we readily see the following:

$$\begin{aligned} W + V_t^P(W, \pi_1, \pi_2) &< \pi_1 + \pi_2 + V_t^A(W, \pi_1, \pi_2), \forall W < W^* \\ W + V_t^P(W, \pi_1, \pi_2) &> \pi_1 + \pi_2 + V_t^A(W, \pi_1, \pi_2), \forall W > W^* \end{aligned} \quad (28)$$

This is precisely the definition of indexability with index $I(t, \pi_1, \pi_2) = W^*$.

From the definition of indexability, we can readily see that uniqueness of $W^*$ is necessary for indexability. □

**Claim 10.** *The ARQ based, single project, active/passive scheduling scheme is indexable and hence the RMAB-v is indexable.*

Our claim is based on extensive simulations, partially reproduced in Fig. 4.

# 6 Structure of the Index Policy

**Proposition 11.** *In any control interval $t$, under indexability, the index function satisfies the following:*

$$\begin{aligned} I(t, \pi_1, \pi_2) &= \pi_1 + \pi_2 \;\; \textit{if} \;\; \pi_1 + \pi_2 \notin (2r, 2p) \\ I(t, \pi_1, \pi_2) &\in [\pi_1 + \pi_2, 2p) \;\; \textit{if} \;\; \pi_1 + \pi_2 \in (2r, 2p) \end{aligned} \quad (29)$$

*Proof.* Under indexability, in any control interval $t$, $W^*$ such that $W + V_t^P(W, \pi_1, \pi_2) = \pi_1 + \pi_2 + V_t^A(W, \pi_1, \pi_2)|_{W=W^*}$ is unique and $I(t, \pi_1, \pi_2) = W^*$. We now prove the first part of the proposition. When $\pi_1 + \pi_2 \geq 2p$, at $W = \pi_1 + \pi_2$, $V_t^A(W, \pi_1, \pi_2) = V_t^P(W, \pi_1, \pi_2)$. This follows from Lemma 8 since $W = \pi_1 + \pi_2 \geq 2p$. Thus $W + V_t^P(W, \pi_1, \pi_2) = \pi_1 + \pi_2 + V_t^A(W, \pi_1, \pi_2)|_{W=\pi_1+\pi_2 \geq 2p}$ leading to $I(t, \pi_1, \pi_2) = W^* = \pi_1 + \pi_2$ when $\pi_1 + \pi_2 \geq 2p$. Using a similar argument for $\pi_1 + \pi_2 \leq 2r$, the first part of the proposition is established. Consider the second part with $2r < \pi_1 + \pi_2 < 2p$. If $W^* \geq 2p$, from Lemma 8, $V_t^A(W, \pi_1, \pi_2) = V_t^P(W, \pi_1, \pi_2)$ leading to $W^* = \pi_1 + \pi_2$. But this contradicts the condition $\pi_1 + \pi_2 < 2p$. Thus $W^* < 2p$. From Lemma 7, $V_t^A(W, \pi_1, \pi_2) - V_t^P(W, \pi_1, \pi_2) \geq 0, \forall W$. Thus $W^* \geq \pi_1 + \pi_2$. This establishes the proposition. □

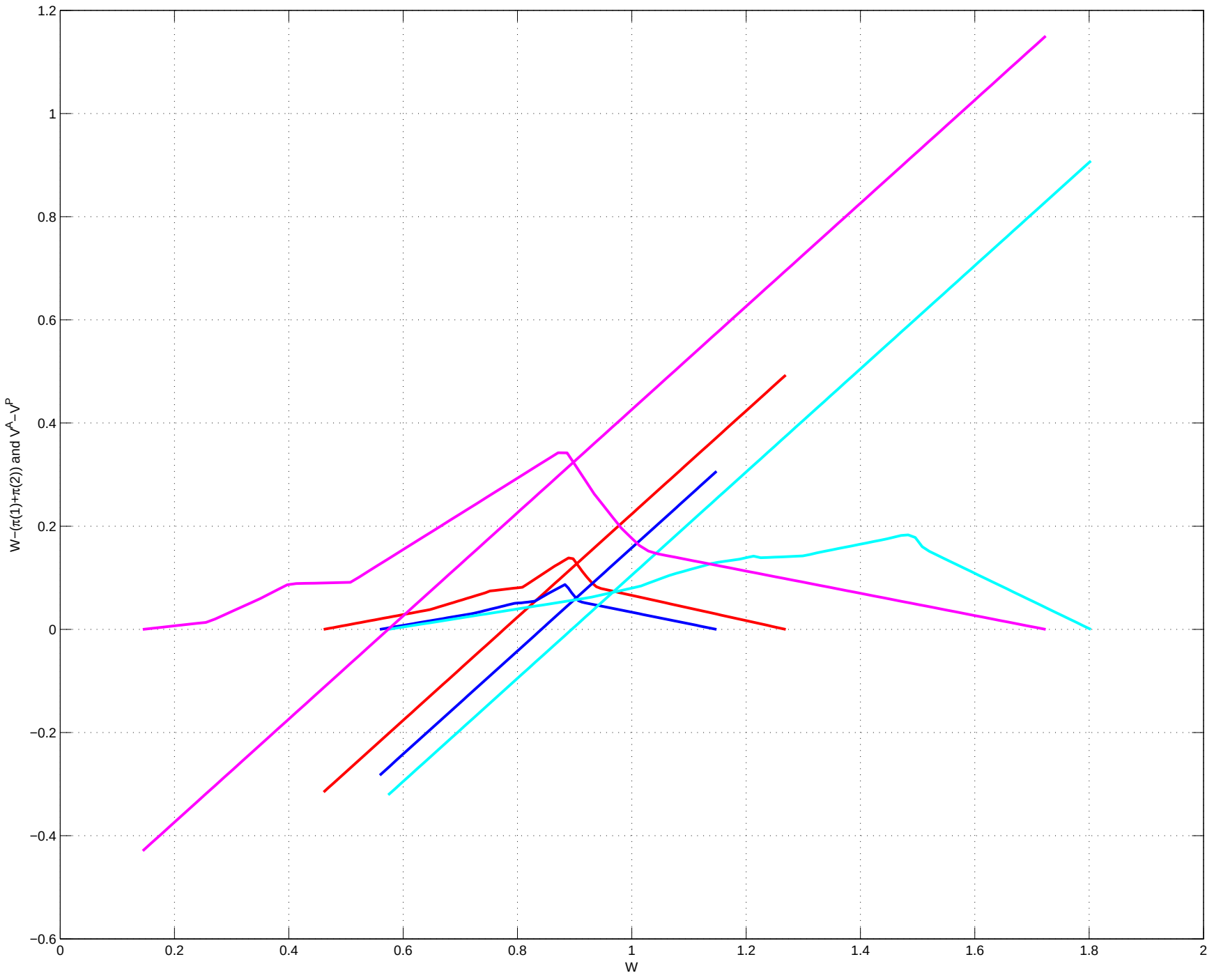

Figure 4: Simulation suggesting indexability of the RMAB-v. For various values of $(\pi_1, \pi_2)$, the function $W - (\pi_1 + \pi_2)$ is shown to intersect the corresponding function $V_t^A(W, \pi_1, \pi_2) - V_t^P(W, \pi_1, \pi_2)$ only once. Index is given by value of $W$ at the point of intersection. Horizon length = 5 is used.

We now show that the index policy partially resembles the greedy policy.

**Proposition 12.** *Under indexability, the index based multi project scheduling policy has the following partial implementation structure: When the ARQ feedback from both the users in the scheduled project are ACKs, reschedule the pair. If both were NACKs, switch to another pair.*

*Proof.* Consider a control interval $t$. Let the state of the $i^{\text{th}}$ project be given by $(\pi_{1,i}, \pi_{2,i})$, $i \in \{1, \dots N\}$. Let $a_t$ be the project scheduled in the control interval $t$. The state of any project $i \neq a_t$, in the next control interval $t-1$ is given by $(T(\pi_{1,i}), T(\pi_{2,i}))$. Since $2r \leq T(\pi_{1,i}) + T(\pi_{2,i}) \leq 2p$, from Proposition 11, the index of project $i$ in control interval $t-1$ is bounded as follows: $2r \leq I_i(t-1, T(\pi_{1,i}), T(\pi_{2,i})) \leq 2p$. The state of the project $a_t$ in the next control interval $\in \{(p,p), (p,r), (r,p), (r,r)\}$ depending upon the nature of the ARQ feedback from the two users constituting the project. Using Proposition 11, upon receiving ACK from both, the index of $a_t$ is $I_{a_t}(t-1, p, p) = 2p \geq I_{i \neq a_t}(t-1, \pi_{1,i}, \pi_{2,i})$ and upon receiving NACK from both $I_{a_t}(t-1, r, r) = 2r \leq I_{i \neq a_t}(t-1, \pi_{1,i}, \pi_{2,i})$. This establishes the partial structure of the index policy. □

**Claim 13.** *If, under indexability, in any control interval $t$, the sum of the belief values of the users in the projects $(\pi_{1,i} + \pi_{2,i})$ are sufficiently separated from each other, the index policy is implemented as the greedy policy in that control interval.*

It was observed from numerical simulations that, with $W$ as the independent variable, $V_t^A(W, \pi_1, \pi_2) - V_t^P(W, \pi_1, \pi_2)$ is closely bounded for all states with equal $\pi_1 + \pi_2$. This led to closely bounded index value $I(t, \pi_1, \pi_2)$ for this family of states identified by $\pi_1 + \pi_2$. In addition, the index value as a function of $\pi_1 + \pi_2$ was observed to have a generally increasing pattern with $\pi_1 + \pi_2$. Together, from these observations, it appears that the index value increases as $\pi_1 + \pi_2$ is increased in sufficiently large steps. This is illustrated in Fig. 5, Fig. 6. Also note that the greedy policy chooses the project with the highest immediate reward, i.e., $\pi_{1,i} + \pi_{2,i}$. This forms the basis for Claim 13. Note that, when every user is scheduled at least once in the past, the belief values of the users are outputs of the functions $T^k(p)$ or $T^k(r)$. Since no two users in a cell are scheduled in the same control interval and taking note of the structure of $T^k$, a natural separation of $\pi_i$ may be guaranteed across users within each cell with the degree of separation being a function of the channel statistics.

# 7 Conclusion

We have addressed the problem of multiuser scheduling with partial channel information in a multi-cell environment. By formulating the scheduling problem jointly with the ARQ based channel learning process and the intercell interference mitigating cell breathing protocol, we obtain optimal joint scheduling policies under various system constraints. We posed the original, unconstrained scheduling problem as a generalized variant of the Restless Multiarmed Bandit processes and introduced the notion of indexability relevant to these processes. We conjectured, with numerical support, that the multicell multiuser scheduling problem is indexable and partially characterized the index policy. Ongoing work focuses on obtaining theoretical support for our conjecture on indexability, in addition to a complete characterization, including performance study, of the index policy.

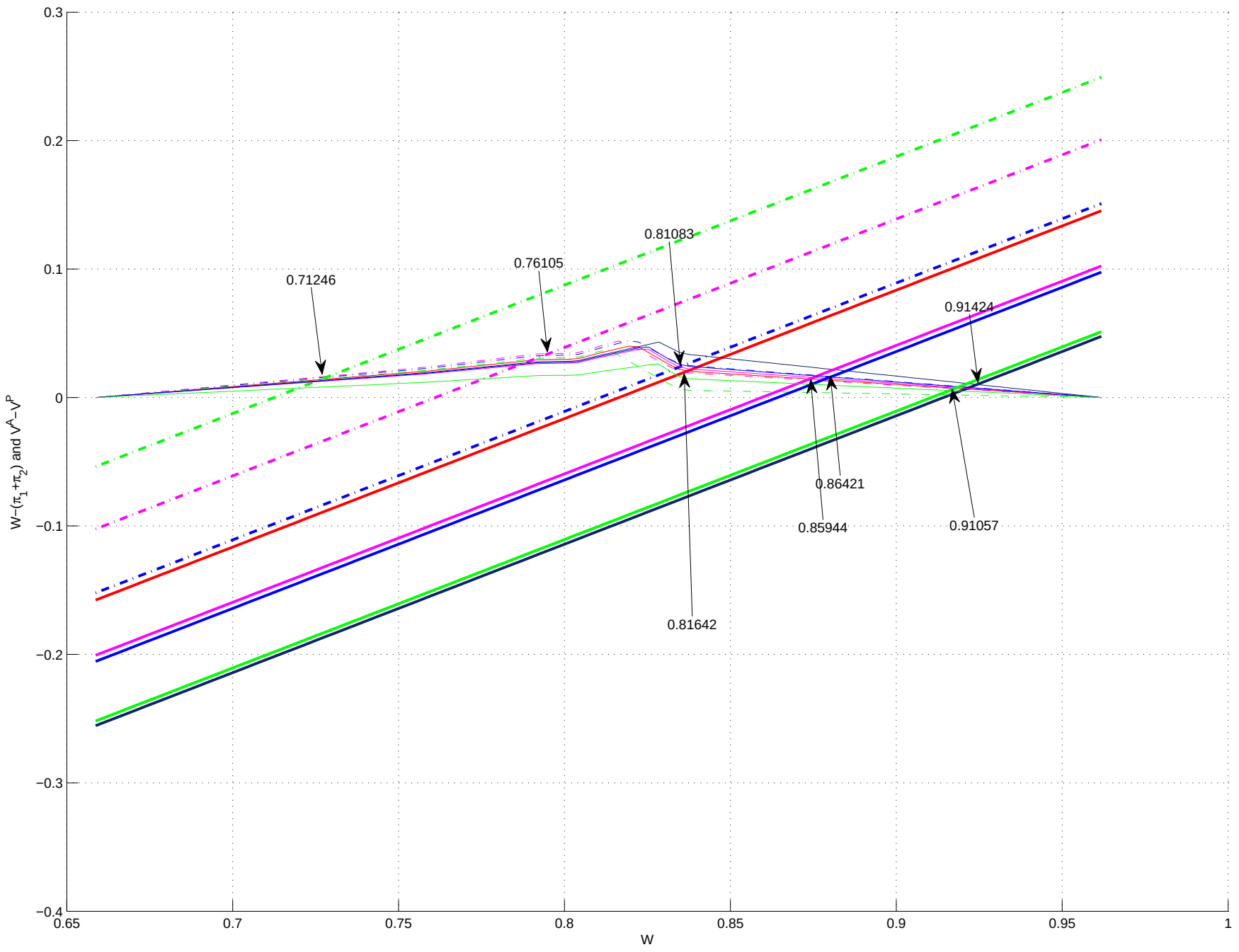

Figure 5: Simulation suggesting index as a monotonic function of $\pi_1 + \pi_2$. $p = 0.4809, r = 0.3294$, horizon size of 5 is used. Intersections of functions $W - (\pi_1 + \pi_2)$ and $V_t^A(W, \pi_1, \pi_2) - V_t^P(W, \pi_1, \pi_2)$ are labeled by $\pi_1 + \pi_2$.

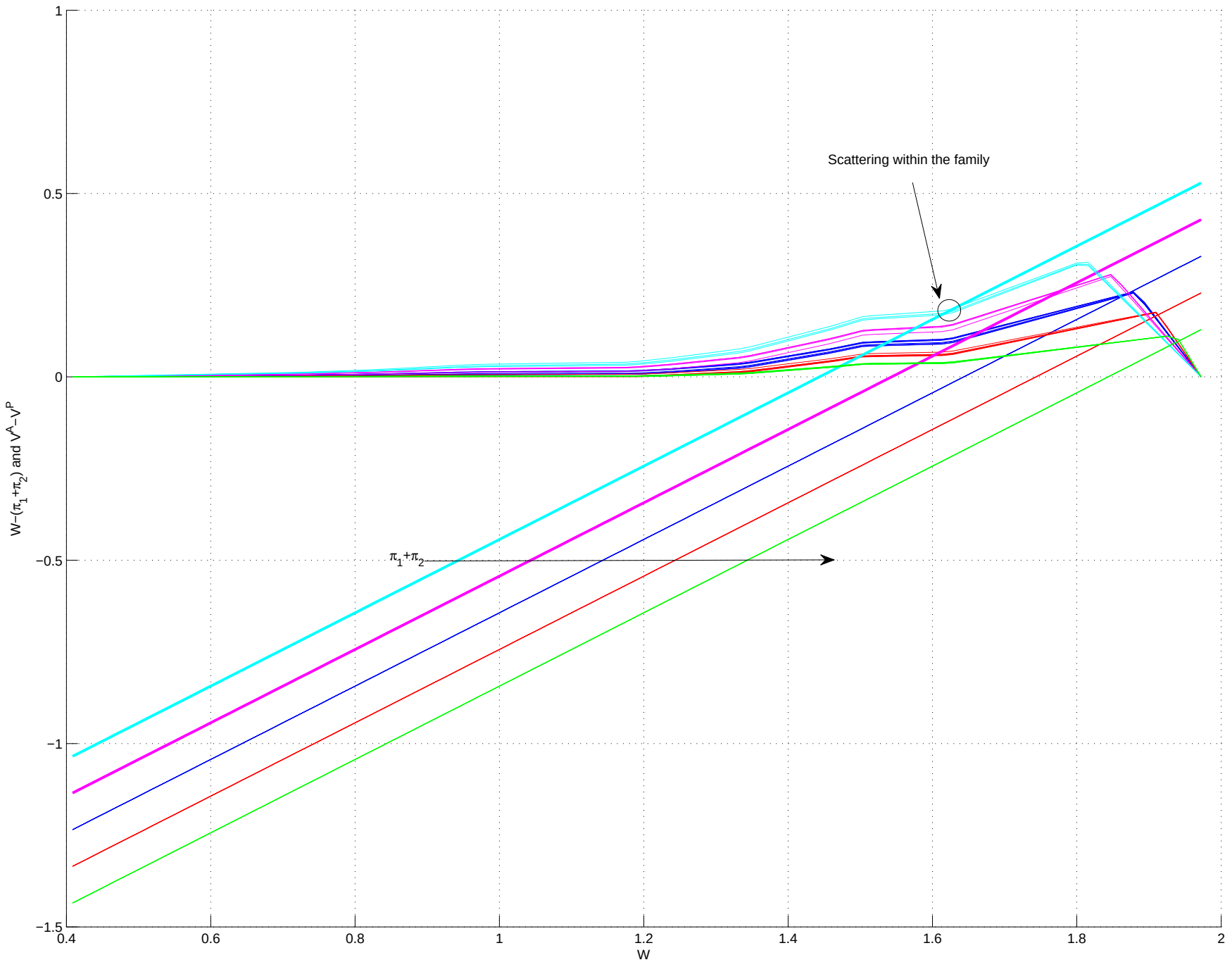

Figure 6: Simulation showing scattering of the index value within the family of states with equal $\pi_1 + \pi_2$. With sufficient spacing between the families $(\pi_1 + \pi_2)$, the monotonicity of index with $\pi_1 + \pi_2$ appears unaffected. $p = 0.9861, r = 0.2043$, horizon size of 5 is used.